'*coexist*': an R package for performing species coexistence modeling and analysis
Youhua Chen
Department of Renewable Resources, University of Alberta, Edmonton, T6G 2H1, Canada
Email: haydi@126.com or youhua@ualberta.ca

ABSTRACT

This letter introduced a new R package *'coexist'* which can perform species coexistence simulation and analysis. The package was initially developed for understanding the role of different combinations of varying species growth rates, dispersal ability, habitat heterogeneity, and competition interaction on influencing species coexistence under fluctuating environment. Further versions could introduce more coexistence models under different ecological conditions. The source package is available at the Google Code (https://code.google.com/p/coexist/) and R CRAN repository (http://cran.r-project.org/web/packages/coexist/).
Keywords: coexistence mechanism, species modeling, fluctuating environment

Understanding biodiversity maintenance mechanisms is the principal objective in ecological studies (Chesson 2000, Bascompte et al. 2006, Salomon et al. 2010). Many theories have been proposed to reveal species coexistence mechanisms, two principal ones were identified as stabilizing and equalizing mechanisms (Chesson 1994, 2000, Chesson and Neuhauser 2002). Stabilizing mechanism supported niche theory while equalizing mechanism advocated for neutral theory.

There were a lot of theoretical work and numerical simulations for explaining species coexistence reasons. However, up to date, there is not a software or program available to perform species coexistence analysis. Also, the role of asymmetric dispersal and differentiation of species' biological traits under fluctuating environments has not been widely understood. To cope with these gaps, I developed an R (R Development Core Team 2013) package called *'coexist'*, which can perform flexible simulations of the outcomes of coexistence of two or multiple species under the conditions that species traits, like intrinsic growth rates, dispersal ability, biotic interaction degree, were varied. The environmental conditions for the species can be varied as well.

Limitation of the current version for the package is that there were many loops involved in the code without optimization. This situation will lead to intensive consumption of computation time, and the pre-configured model with two species and simulation time of 10000 steps will lead to 3 hours. However, if the readers just want to learn how to perform individual-based species coexistence simulation and modeling, the package should be valuable.

The resultant simulation outcomes could be read, saved and opened using the tailored functions in the package. Typically the files would be saved with a postfix "00yh.dat", which will be served as the sign for batch reading of all the files inside a folder. Therefore, if the users have many models for simulations, it is quite convenient to analyze these models simultaneously using the batch-handling commands in the packages.

Currently, the package also integrated several graphical functions as well. Coexistence contour plots for the combination of each pair of parameters can be generated in the form of heatmaps (different colors in the grids indicated different coexistence probabilities).

Figure 1 showed a heatmap for illustrating the outcomes generated from the linear combination of two parameters' value spaces: growth rates for two species (r1 and r2). The result was the outcome generated by the Model 1 in the following text.

DEMONSTRATION

Here is the procedure to generate a model configuration and perform analysis.

Firstly, in the package there is a simple function *sampletype* to help generate model names (in string form) with corresponding numbers (thus it should be easier without writing model names repeatedly). I can script the spatial pattern names as well.

```
#spatial patterns of a a single parameter in the simulations
#1=decrease,2=increase,3=constant,4=mosaiclow,5=mosaichigh
modelnum=5
h<-vector()
h[1]<-"decrease"
h[2]<-"increase"
h[3]<-"constant"
h[4]<-"mosaiclow"
h[5]<-"mosaichigh"
```

It is noted that in the packages only spatial patterns were provided: "decrease" indicated that the parameter value is decreased from source to sink patches, while "increase" is opposite. "constant" demonstrated that the parameter is not influenced by the patch ordering. Last, "mosaiclow" is the one-by-one switching of the values (one of two values, either the high or low one, will be assigned to the parameter) between neighboring patches, but the first value assigned to the parameter (in the second patch adjacent to the source) is set to low. "mosaichigh" is almost the same, but the first value occurred in the second patch is set to the high one.

If I set a combination of spatial patterns for each parameter as,

*t<-c(h[1],h[4],h[3],h[3],h[1],h[3],h[3])*

After defining a sampled parameter value space as,

*parspace<-c(.1,.25,.5,.75,.9) #serve as the searching points for each parameter in the interval [0-1]*

Then it is ready to use for performing simulation. Before doing that, I should take a view about the functions for simulations. There are several versions of simulation functions.

*sim.coarse()* is for 2-species model. If I assume that both species' dispersal abilities were not different and marked by the iconicity of neighboring patches, then I have the following 7 paramters in order: habitat quality/preference for species 1 and 2, species growth rates for 1 and 2, patch connectivity, competition coefficients for species 1 and 2.

*fast.flexsim()* is for multiple species model (species number>=2), each species will have 4 parameters, because I considered that connectivity is not that important as species dispersal capabilities. Therefore, the parameter of habitat connectivity was replaced by two parameters for measuring dispersal ability of species 1 and 2 respectively. Therefore, in this flexible simulation, I would have parameter number=4*species number.

Moreover, this function can relax the assumption that the source supply of species is constant across the simulation time. Instead, there three source supply patterns are available, which was generated from another function *flex.dispersal()*: "*cochange*" means that all species will have identical source supply for each time step; "*flexible*" means that all species will have their own source supply at each time step, but all followed a same normal distribution; "*constant*" means that the species supply from the source is unchanged across time steps and identical among species.

*flexsim()* identical to the function *fast.flexsim()*, but the latter should be faster a little bit (but remember that there are so many loops in the package, all the simulations are not fast at all!).

*sim.refinement()*: if I am interested into exploring further about a specific pair parameter space while setting others unchanged. This function can help do the refinement for searching the parameters' values at a smaller interval. It is only for 7-parameter 2-species model, therefore is limited to use with *sim.coarse()*.

Now I can run the model with my configuration of spatial patterns for each parameter.

*Model1<- sim.coarse(scale=2,allee=1,T=500,prange=parspace,type=t,path="c://test/out1")*

In the function, parameter *allee* indicated the allee effect (how small the population will be wiped out). *scale* is scale factor used to weigh connectivity parameter (making species colonization rates smaller). *type* is the model configuration as above. *path* is the local disk path to save the simulation output data.

If I run multiple species model, I should firstly modify the spatial pattern vector to be the length of 2*species number and keep in mind that the parameters always follow the order as *(habitat quality for sp1,sp2…;growth rates for sp1,sp2…;dispersal rates for sp1,sp2…;competition rates for sp1,sp2…)*!

For example, if I still have 2 species (more species are OK, but the simulation time will be incredibly huge under R environment).

*t<-c(h[2],h[2],h[3],h[3],h[3],h[3],h[3],h[3]) #indicate a simple model that habitat quality for the two species are better from source to sink patches*

So, the model becomes,

*Model2<-fast.flexsim(scale=2,allee=1,T=400,parcomb=allcomb,spnum=2,sourcetype="constant",type=t,path="c://test/out2")*

The parameters in *fast.flexsim* function are a bit different, which "*parcomb*" replaced "*prange*". Because the the combinations of all parameters are calculated firstly prior to simulation, with the purpose to reduce simulation budget.

So, "parcomb" accepts a matrix, not a vector, which is the combination of parameter spaces. The generation of the matrix can use a function called make.parcomb() as follows,

*parnum=8*

*parv<-c(.1,.9) #paramter value space for a single paramter*

*allcomb<-make.parcomb(parv,parnum,path="c://outcome/index.dat")*

*dim(allcomb)[1] #=256*

It will have $2^8$ combinations=256.

After simulation, I can plot the density plot as a heatmap for exploring the influence of pairwise parameter combination on influencing species coexistence.

The function *batch.mpaircomp()* can generate the mean and variance of density matrix for each pair of parameters. If I want to look at species coexistence status only varying one parameter, the function *batch.monepar()* can be used.

Plotting heatmaps and making pdf graphics have the functions *batch.pdf.pairpar()* and *batch.pdf.onepar()* respectively for pairwise parameter and single parameter cases.

For example, for the Model 2, I can have,

*pairparlist<-batch.mpaircomp(Model2,coenum=2,spnum=spnum,parameters=**parv**)*

*singleparlist<-batch.monepar(Model2,coenum=2,island,spnum=spnum,parameters= **parv**)*
*batch.pdf.pairpar(pairparlist,pagesetup=c(2,2),path="c://test/pairpars.coexist")*
*batch.pdf.onepar(singleparlist,pagesetup=c(2,2),path="c://test/onepar.coexist")*

Finally, by packing all the codes together, I can perform simulation of species coexistence under different parameter setting as follows,

```
#spatial patterns of a a single parameter in the simulations
#1=decrease,2=increase,3=constant,4=mosaiclow,5=mosaichigh
library(coexist)
h<-vector()
h[1]<-"decrease"
h[2]<-"increase"
h[3]<-"constant"
h[4]<-"mosaiclow"
h[5]<-"mosaichigh"

#two species model without dispersal difference
t<-c(h[1],h[4],h[3],h[3],h[1],h[3],h[3])
parspace<-c(.1,.25,.5,.75,.9) #serve as the searching points for each parameter in the interval [0-1]
Model1<- sim.coarse(scale=2,allee=1,T=1000,prange=parspace,type=t,path="c://test/out1")

#multiple species model
parnum=8
parv<-c(.1,.9) #paramter value space for a single paramter
allcomb<-make.parcomb(parv,parnum,path="c://outcome/index.dat")
t<-c(h[2],h[2],h[3],h[3],h[3],h[3],h[3],h[3]) #indicate a simple model that habitat quality for the two species are better from source to sink patches
Model2<-fast.flexsim(scale=2,allee=1,T=300,parcomb=allcomb,spnum=2,sourcetype="constant",type=t,path="c://test/out2") #if T>1000, very slow

#heatmap plotting and saving
pairparlist<-batch.mpaircomp(Model2,coenum=2,spnum=spnum,parameters=parv)
singleparlist<-batch.monepar(Model2,coenum=2,island,spnum=spnum,parameters= parv)
batch.pdf.pairpar(pairparlist,pagesetup=c(2,2),path="c://test/pairpars.coexist")
batch.pdf.onepar(singleparlist,pagesetup=c(2,2),path="c://test/onepar.coexist")
```

References

Bascompte, J., P. Jordano, and J. Olesen. 2006. Asymmetric coevolutionary netowrks facilitate biodiversity maintenance. Science 312:431–433.

Chesson, P. 1994. Multispecies competition in variable environments. Theoretical Population Biology 45:227–276.

Chesson, P. 2000. Mechanisms of maintence of species diversity. Annual Review of Ecology and Systematics 31:343–366.

Chesson, P., and C. Neuhauser. 2002. Intraspecific aggregation and species coexistence. Trends in ecology & evolution 17:210–211.

R Development Core Team. 2013. R: A Language and Environment for Statistical Computing, Vienna, Austria. ISBN 3-900051-07-0, URL http://www.R-project.org. R Foundation for Statistical Computing, Vienna, Austria.

Salomon, Y., S. R. Connolly, and L. Bode. 2010. Effects of asymmetric dispersal on the coexistence of competing species. Ecology letters 13:432–41.

Figure 1. Coexistence probability by partialling out the influence of other parameters but keeping the effect from the parameter pair combination of the varying growth rates of two species model. Darker color indicated higher coexistence probability. Necessary model configuration for this simulation in "coexist" package was (1) habitat quality for species 1 is decreased from source to sink patches; while for species 2 is always in transition across the patches; (2) connectivity between the neighboring patches was decreased from the source to sink. It was derived from the Model 1.

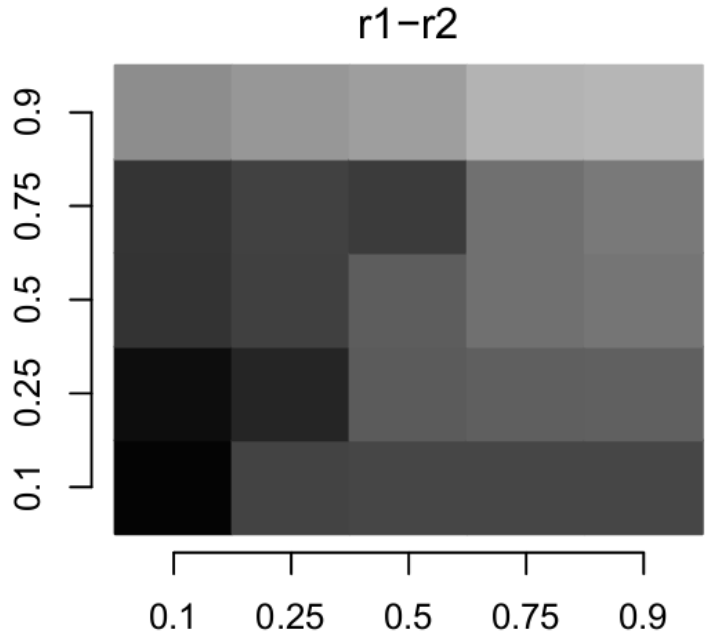